\documentclass[11pt]{article}
\usepackage{amsfonts}
\usepackage{amssymb}
\usepackage{amsmath}
\usepackage{latexsym}
\usepackage{graphicx}
\usepackage{mathrsfs}
\usepackage[english]{babel}
\usepackage[usenames]{color}
\input epsf

\renewcommand{\theequation}{\arabic{section}.\arabic{equation}}

\newcommand\be{\begin{equation}}
\newcommand\bea{\begin{eqnarray}}
\newcommand\ee{\end{equation}}
\newcommand\eea{\end{eqnarray}}
\newcommand\h{\frac{1}{2}}
\newcommand\Regge{\alpha'}
\newcommand{\bdm}{\begin{displaymath}}
\newcommand{\edm}{\end{displaymath}}

\newcommand{\mathsym}[1]{{}}

\newcommand{\f}[2]{\frac{#1}{#2}}
\newcommand{\p}{\phantom{a}}
\newcommand{\bref}[1]{(\ref{#1})}

\begin{document}

\begin{flushright}
%OHSTPY-HEP-T-03-012\\
\end{flushright}
\vspace{20mm}
\begin{center}
{\LARGE  Pair creation in non-extremal fuzzball geometries}\\
\vspace{18mm}
{\bf Borun D. Chowdhury}$^{a,}$\footnote{E-mail: {\tt borundev@mps.ohio-state.edu}.}, 
{\bf and Samir D. Mathur}$^{a,}$\footnote{E-mail: {\tt mathur@mps.ohio-state.edu}.}\\
\vspace{8mm}
$^{a}$Department of Physics,\\ The Ohio State University,\\ Columbus,
Ohio, USA 43210\\
\vspace{4mm}
\end{center}
\vspace{10mm}
\thispagestyle{empty}
\begin{abstract}

It is possible to construct a special family of nonextremal black hole microstates. These microstates are unstable, and emit radiation at a rate which is found to exactly equal the Hawking radiation rate predicted for them by the dual CFT. In this paper we analyze in more detail the nature of the radiation created by these unstable modes.  The energy and angular momentum of the mode is found to be localized in two regions: one near infinity corresponding to the emitted quanta, and the other in the ergoregion which is deep inside the interior of the geometry. The energy and angular momenta are equal and opposite for these two contributions, as expected for emission from ergoregions. We conjecture that more general nonextremal microstates will possess  ergoregions (with no axial symmetry), and radiation from these regions can be part of the general Hawking emission for the microstates.

\end{abstract}
\newpage
\setcounter{page}{1}

\section{Introduction}\label{intr}
\setcounter{equation}{0}

In the traditional picture of a black hole, the region near the horizon is in a vacuum state. Semiclassical evolution of quantum fields on this background produces particle-antiparticle pairs. One member of this pair falls into the hole and reduces its mass, while the other member escapes to infinity as Hawking radiation \cite{hawkingone}.

The problem with this picture is that we get information loss; the escaping quanta have no information about the matter which made the hole \cite{hawkingtwo}. To solve the information paradox we have to see what can change in this picture, and how the escaping quanta carry the information of the state.

A black hole has a large number $e^{S_{bek}}$ of states, where $S_{bek}$ is the Bekenstein entropy of the hole. We now have some understanding of the interior structure of the hole: the information of the matter inside is spread throughout the interior, making a fuzzball \cite{fuzzball1,fuzzball2}. For very special microstates the fuzzball can be a classical geometry rather than a `quantum fuzz', and it is useful to first analyze the behavior of such special states. 

In \cite{ross} a family of  nonextremal microstates were constructed. In \cite{myers} it was found that these geometries were unstable to radiation of a scalar field. Finally, in \cite{chowdhurymathur} it was shown that this radiation is just the `Hawking radiation' that would be expected from this particular microstate. More precisely, one looks at the microscopic emission process from states of the dual CFT. It is known that if we take a microstate with a generic excitation structure, then we reproduce the gross properties of Hawking radiation from the corresponding hole \cite{radiation}.\footnote{Recently the gross properties of superradiance from rotating black holes were also reproduced in a similar fashion \cite{emparan}.}   Performing the same computation with  the special CFT microstate  gives exactly the radiation found from the unstable geometry. 
Thus we have here a simple example where we can see explicitly the `Hawking radiation' emerging from a particular (rather nongeneric) microstate. 

In this paper we study the nature of this radiation in more detail. We find the following. The scalar field waveform can be split, to a good approximation,  into two parts.
One part escapes to infinity, and we compute the energy $\omega$ and angular momentum $j$ carried by such quanta. The other part settles deep into the `cap region' of the geometry; the quanta here have energy $-\omega$ and angular momenta $-j$. Thus these two halves of the waveform correspond to particle-antiparticle pairs. The waveform that collects in the `cap' region is seen to be localized in the ergoregion. 
These properties are  expected from particle production in an ergoregion \cite{kang}, but it is interesting to see the explicit construction of the two halves of the wavefunction because very few examples of ergoregions without horizons have been studied (for examples, see \cite{kang,friedman,cominsschutz}).  

We then discuss the nature of radiation from nonextremal fuzzballs.
We  conjecture a picture of ergoregion emission that would apply to a large class of fuzzball geometries. These geometries would have no axial symmetry or net rotation, but the essential feature they share with the microstate of \cite{ross} is that there is no Killing vector that is timelike everywhere. This leads to the absence of a time independent vacuum state, and there will be particle production in general.

We close with a general discussion of some properties of fuzzballs.

\section{The microstate geometry}
\setcounter{equation}{0}

Let us start by recalling the microstate geometries that we will consider. These geometries were constructed in \cite{ross}. We describe the geometries, and then explain the limits of parameters that we will take in carrying out our computations. 

\subsection{The supergravity solution}

Let us recall the setting for the geometries of \cite{ross}. One starts with supergravity solutions for arbitrary charges, rotation and angular momenta \cite{cveticyoum,cveticlarsen}, and then chooses parameters such that in the dual CFT there is a unique state with those quantum numbers. This procedure is found to give a smooth geometry without a horizon, which represents the given CFT microstate. This process was used for 2-charge extremal geometries in \cite{balmm}, for 3-charge extremal geometries in \cite{gms}, and for constructing a family of nonextremal geometries in \cite{ross}. It is these nonextremal geometries that we will use in the present paper.

Take type IIB string theory, and compactify 10-dimensional spacetime as
\be
M_{9,1}\rightarrow M_{4,1}\times T^4\times S^1
\ee
The volume of $T^4$ is $(2\pi)^4 V$ and the length of $S^1$ is $(2\pi) R$. The $T^4$ is described by coordinates $z_i$ and the $S^1$ by a coordinate $y$. The noncompact $M_{4,1}$ is described by a time coordinate $t$, a radial coordinate $r$, and  angular $S^3$  coordinates $\theta, \psi, \phi$. The solution will have angular momenta along $\psi, \phi$, called $J_\psi, J_\phi$, captured by two parameters $a_1, a_2$.
The solutions will carry three kinds of charges. We have $n_1$ units of D1 charge along $S^1$, $n_5$ units of D5 charge wrapped on $T^4\times S^1$, and $n_p$ units of momentum charge P along $S^1$. 
These charges  will be described in the solution by three parameters $\delta_1, \delta_5, \delta_p$. 

In this paper we will look at  states where the P charge is zero ($\delta_i=n_p=0$), since this case will suffice to bring out the observations that we wish to make. It turns out that $n_p=0$ implies that one of the angular momenta vanish: $J_\phi=0$. The resulting geometries are (in the string frame)
\begin{eqnarray} \label{2charge}
ds^2&=&-\frac{f-M}{\sqrt{\tilde{H}_{1} \tilde{H}_{5}}}
dt^2+\frac{f}{\sqrt{\tilde{H}_{1} \tilde{H}_{5}}}dy^2+\sqrt{\tilde{H}_{1} \tilde{H}_{5}}
\left(\frac{ dr^2}{ r^2+a_{1}^2 - M}
+d\theta^2 \right)\nonumber \\ 
&&+\left( \sqrt{\tilde{H}_{1}
\tilde{H}_{5}} + a_1^2 \frac{( \tilde{H}_{1} + \tilde{H}_{5}
-f+M) \cos^2\theta}{\sqrt{\tilde{H}_{1} \tilde{H}_{5}}}  \right) \cos^2
\theta d \psi^2 \nonumber \\ 
&& +\left( \sqrt{\tilde{H}_{1}
\tilde{H}_{5}} -a_1^2 \frac{(\tilde{H}_{1} + \tilde{H}_{5}
-f) \sin^2\theta}{\sqrt{\tilde{H}_{1} \tilde{H}_{5}}}\right) \sin^2
\theta d \phi^2 \nonumber \\ 
&&+ \frac{2M \cos^2 \theta}{\sqrt{\tilde{H}_{1} \tilde{H}_{5}}}(a_1
c_1 c_5 ) dt  d\psi +\frac{2M \sin^2 \theta}{\sqrt{\tilde{H}_{1} \tilde{H}_{5}}}(a_1
s_1 s_5 ) dyd\phi + \sqrt{\frac{\tilde{H}_1}{\tilde{H}_5}}\sum_{i=1}^4
dz_i^2 \label{Eqn:metric}
\end{eqnarray}
where
\be
c_i = \cosh \delta_i, \quad s_i=\sinh \delta_i
\ee
\begin{eqnarray} 
\tilde{H}_{i}=f+M\sinh^2\delta_i, \quad
f=r^2+a_1^2\sin^2\theta, \label{Def:FandH}
\end{eqnarray}
The D1 and D5 charges of the solution produce a RR 2-form gauge field. The RR 2-form gauge field and the dilaton are given in  \cite{ross,gms}. The angular momenta are given by
\bea
J_\psi &=& -  \f{\pi M}{4 G^{(5)}} a_1 c_1 c_5   \\
J_\phi &=& 0
\eea
It is convenient to define
\be 
Q_1=M\sinh\delta_1\cosh\delta_1, ~~Q_5=M\sinh\delta_5\cosh\delta_5
\label{qdef}
\ee
The integer charges of the solution are related to the $Q_i$ through
\bea
Q_1&=& \frac{g \Regge^3}{V} n_1 \nonumber \\ 
Q_5 &=& g \Regge n_5  \label{q1q5}
\eea
We must further choose
\be
\f{M s_1 s_5}{\sqrt{a_1^2-M}}=R, \qquad  \f{a_1}{\sqrt{a_1^2 - M}} =m\label{Eqn:Smooth}
\ee
where $R$ is the radius of the $S^1$ and $m \in \mathbb Z$. The geometries are then regular solutions representing microstates of the non-extremal D1-D5 system \cite{ross}.

In our work below we will ignore the torus $T^4$ since none of our variables depend on the torus coordinates. We will work with the 6-d Einstein metric unless otherwise mentioned. It turns out that this metric is the same as (\ref{Eqn:metric}) with the torus contribution discarded.

\subsection{The large R limit}

As explained in \cite{chowdhurymathur}, if we want to relate our computations to a dual CFT description then we need to have a large AdS type region in our geometry. Such a region is obtained if we let the radius $R$ of the $S^1$ be large. 
The large R limit is defined by
\be
\epsilon \equiv \f{\sqrt{Q_1Q_5}}{R^2} \ll  1 \label{Def:epsilon}
\ee
We would now like to express the parameters of the solution (\ref{Eqn:metric}) in a way that manifests their behavior in this large $R$ limit. 
From \bref{Eqn:Smooth} we get
\be
R^2=(m^2-1) M s_1^2 s_5^2
\ee
From \bref{q1q5}  we can see that $Q_1,Q_5$ do not depend on $R$. In this large $R$ limit we will have $M\ll Q$ and 
\be
s_1 \approx c_1, \quad s_5\approx c_5
\label{firsteq}
\ee
which gives with \bref{qdef}
\be
M s_1^2 =Q_1, \qquad Ms_5^2 = Q_5
\ee
We will assume that $Q_1$ and $Q_5$ are of the same order.

We see that 
\be
M=(m^2-1) \f{Q_1 Q_5}{R^2}
\ee
and we get
\be
a_1 = m \f{\sqrt{Q_1 Q_5}}{R} 
\ee
With these expressions for $M$ and $a_1$  we get from \bref{Def:FandH}
\bea
f&=&r^2 + m^2 \f{Q_1Q_5}{R^2} \sin^2 \theta \nonumber \\
\tilde{H}_i &=& r^2+ Q_i \label{Eqn:FandHLargeR}
\eea

\subsection{The inner and outer regions}

In the large $R$ limit we can separate the geometry into two regions: an `inner region' which is an $AdS$ type geometry and an outer region which is essentially flat space. These two regions are connected by a region around $r\sim (Q_1Q_5)^{\f{1}{4}}$ which we will call the `neck'. 

\subsubsection{Inner Region: $r^2  \ll \sqrt{Q_1 Q_5}$}

In this region 
\bea
f&=&r^2 + m^2  \f{Q_1Q_5}{R^2} \sin^2 \theta \nonumber \\
\tilde{H}_i &=& Q_i 
\eea
The metric takes a simple form in terms of the coordinates
\bea
\tau \equiv \f{t}{R}, \qquad \varphi \equiv \f{y}{R}, \qquad \rho \equiv \f{r R}{\sqrt{Q_1 Q_5}} \label{Def:AdSCood}
\eea
In these coordinates the  metric in the inner region is
\bea
ds^2 &=& \sqrt{Q_1 Q_5} \Big[ \left (-(1+ \rho^2) d\tau^2 + \f{d \rho^2}{(1+ \rho^2)} + \rho^2 d\varphi^2 \right ) \nonumber \\
&& + \left ( d \theta^2 + \cos^2 \theta ( d \psi + m d \tau)^2 + \sin^2 \theta (d \phi + m d\varphi)^2 \right) \Big]
\label{metricinner}
\eea
This geometry has the form of $AdS_3$ with an $S^3$ fibred over the $AdS_3$.  The fibration is characterized by the integer $m$. The $AdS_3$ and the $S^3$ each have curvature radius $(Q_1 Q_5)^\f{1}{4}$. 

The condition defining the inner region  $0<r\ll  (Q_1 Q_5)^\f{1}{4}$ is equivalent to $0< \rho\ll  \f{R}{(Q_1 Q_5)^\f{1}{4}}$. In the large $R$ limit the radial coordinate of the $AdS$ region extends over `many many curvature radii' before we reach the `neck'. Thus we have a large $AdS$ region and a good description in terms of a dual CFT.

\subsubsection{The Outer Region: $r^2 \gg \sqrt{Q_1Q_5}$}

For our purposes it will be adequate to approximate the metric in this region by its leading approximation which is flat spacetime:
\be
ds^2 = - dt^2 + dy^2 + dr^2 + r^2 d\Omega_3^2 
\label{flat}
\ee

\section{The Minimally coupled scalar}
\setcounter{equation}{0}

The geometries considered above are nonextremal, and have an instability that leads to energy being radiated to infinity \cite{myers}. 
In this section we recall the wave-equation satisfied by this scalar. The solution to this wave-equation is reproduced in appendix \ref{aone}. From this solution we will extract the form of the wavefunction in the inner and outer regions; these two parts will correspond to the two members of particle pairs created in the ergoregion of the geometry.

We consider a minimally coupled scalar field in the 6-dimensional geometry obtained by dimensional reduction on the $T^4$.  Such a scalar arises for instance from $h_{ij}$, which is the graviton with both indices along the 
$T^4$. The wave equation for the scalar is
\be
\Box \Psi =0
\ee
We can separate variables with the ansatz \cite{ross,myers,cveticlarsen}
\be
\Psi = exp(- i \omega t + i \lambda \f{y}{R} + i m_\psi \psi + i m_\phi \phi) \chi(\theta) h(r)
\ee
We will set $m_\phi=\lambda=0$, so we will have
\be
\Psi = e^{-i (\omega t - m_\psi \psi ) }\chi(\theta) h(r) \label{Eqn:Ansatz}
\ee

The wave equation is solved separately in the inner and outer regions, and these solutions are then matched in their domain of overlap.

We look for solutions which are regular in the interior and purely `outgoing' at infinity. Such solutions were derived in  \cite{myers} and are rederived in the large R limit in appendix \ref{aone}. From eqs. \bref{Eqn:WavefunctionInAppendix}, \bref{Eqn:WavefunctionOutAppendix} we see that the wavefunction in the inner and outer regions is given respectively by
 \bea
\Psi_{in} &=& e^{ \omega_I t} e^{-i(\omega_R t - m_\psi \psi) }\chi(\theta) (1+x)^{-(\f{l+2}{2})}   \label{Eqn:WavefunctionIn} \\
\Psi_{out} &=&   (i)^{3l}  e^{- i \f{\pi}{4}} \sqrt{ \f{Q_1 Q_5}{R} } \sqrt{ \f{\omega_I}{\omega_R}}   \chi(\theta) \f{1}{r^\f{3}{2}}  e^{ \omega_I (t-r)} e^{-i(\omega_R( t-r) - m_\psi \psi) }  \label{Eqn:WavefunctionOut}
\eea
where
\be
x \equiv \f{r^2 R^2}{Q_1 Q_5} \label{Def:xBulk}
\ee
For this solution the real part of the frequency is (eq. \bref{Eqn:OmegaReal}) 
\be
\omega_R= \f{1}{R}(-l -2 - m_\psi m)
\label{wtwo}
\ee
The imaginary part of $\omega$ is (eq. \bref{Eqn:OmegaImag} using \bref{Def:Kappa}) 
\be
\omega_I = \f{1}{R} \f{2 \pi}{(l!)^2} \left( \f{\omega_R^2 Q_1 Q_5}{4R^2} \right)^{(l+1)}
\ee

Note that $\omega_I>0$, so the solution grows exponentially in time. Thus there is an instability in this geometry which leads to the creation of quanta of the scalar field. This particle creation can be traced back to the existence of an ergoregion in the geometry \cite{myers}.  We would like to see in more detail where these created particles are accumulating, and what quantum numbers they carry.

Note that the solutions (\ref{Eqn:WavefunctionIn}),(\ref{Eqn:WavefunctionOut}) are written in the limit of large $R$, and so there will be corrections due to the fact that $R$ is not strictly infinite. In appendix (\ref{compare}) we show that  the correction term in the inner region is small compared to the leading order expression  (\ref{Eqn:WavefunctionIn}); a similar check can be carried out for the outer region.

\section{Conserved Charges}
\setcounter{equation}{0}

In the above section we separated the scalar wavefunction into two parts: one flowing off to infinity, and one localized in the inner region of the geometry. In this section we will compute the conserved charges associated to these two parts of the wavefunction. The wavefunction grows exponentially in time, so the total value of any conserved charge should vanish for the wavefunction. But the two halves of the wavefunction will have nonzero values for these charges, which should be equal and opposite. We compute these values, and check that they indeed are equal and opposite for the two halves of the wavefunction.
In appendix \ref{athree} we show that the wavefunction in the inner region is localized in the ergoregion, so  we verify the  picture of pair creation in ergoregions: particle pairs are produced, one member settles down in the ergoregion, while the other flows off to infinity.

There are four Killing vectors for our geometry -  the translations $\xi^{\mu}_{(\phi)} = \delta^{\mu}_\phi$,~ $\xi^{\mu}_{(\psi)} = \delta^{\mu}_\psi,~~\xi^{\mu}_{(y)} = \delta^\mu_y,~~\xi^\mu_{(t)} = \delta^\mu_t$. 
The geometry itself has no rotation in the $\phi$ direction since $J_\phi=0$, and no momentum in the $y$ direction since the momentum charge
$n_p=0$. Since we have taken $m_\phi=\lambda=0$ the perturbation will also not carry these charges. Then the nontrivial conserved quantities are given by
\be
H = - \int \xi^\mu_{(t)} T_\mu^{\p \nu} dS_\nu =- \int T_t^{\p \nu} dS_\nu 
\ee
and
\be
L = \int \xi^\mu_{(\psi)} T_\mu^{\p \nu} dS_\nu = \int T_\psi^{\p \nu} dS_\nu
\ee
where $T^\mu{}_\nu$ is the energy-momentum tensor of the scalar field and the integral extends over a spacelike hypersurface with volume element $dS_\mu$. $H$ measures energy, while $L$ measures angular momentum in the $\psi$ direction.

A simple spacelike surface that we can take is the surface $t=constant$. For this surface the normal $n_\mu = \partial_\mu (t) = \delta^t_\mu$ has the norm $g^{tt}$. We have \cite{ross}
\be
g^{tt}= - \f{1}{\sqrt{\tilde H_1 \tilde  H_5}} \left( f + M +  M s_1^2 + M s_5^2 + \f{M^2 c_1^2 c_5^2 }{r^2+a_1^2-M} \right)
\ee
It can be shown that $g^{tt}<0$ for the regular solutions of \cite{ross}. For our present purposes we are interested in the large $R$ limit, so let us write $g^{tt}$ in this limit, using (\ref{firsteq})-(\ref{Eqn:FandHLargeR}):
\be
g^{tt} = -\f{1}{ \sqrt{(r^2+ Q_1)(r^2 + Q_5)}} \left(r^2 + Q_1 +Q_5 + \f{Q_1 Q_5}{r^2+ \f{Q_1 Q_5}{R^2}} \right)
\ee
which is manifestly negative. 
Thus the $t=constant$ surface is indeed spacelike everywhere. With this surface we have
\newcommand{\angles}{\mathcal A}
\be
dS_\mu = \delta^t_\mu \sqrt{-g} dr d\angles
\ee
where we define $d \angles=d \theta d\phi d \psi dy$. We have
\bea
H &=& -\int \sqrt{-g} dr d \angles ~ T_t^{\p t} \nonumber \\
L  &=& \int  \sqrt{-g} dr d\angles  ~ T_\psi^{\p t} 
\eea
The minimally coupled scalar wave equation comes from the action
\be
S = \int d^d x \sqrt{-g} \partial_\mu \Psi^* \partial^\mu \Psi
\ee
The energy-momentum tensor is 
\be
T_{\mu \nu}  = \f{2}{\sqrt{-g}} \f{\delta S}{\delta g^{\mu \nu}} =\partial_\mu \Psi \partial_\nu \Psi^* +\partial_\nu \Psi \partial_\mu \Psi^* - g_{\mu \nu} (\partial_\mu \Psi \partial^\mu \Psi^*)
\ee
We have, using (\ref{Eqn:Ansatz})
\bea
T_{\psi}^{\p t}&=&\partial_\psi \Psi \partial^t \Psi^* + \partial_\psi \Psi^* \partial^t \Psi  \nonumber \\
&=&(i m_\psi) \Psi \left( g^{tt} \partial_t \Psi^*+g^{t \psi} \partial_\psi \Psi^* \right) + (-i m_\psi) \Psi^* \left( g^{tt} \partial_t \Psi +g^{t \psi} \partial_\psi \Psi \right) \nonumber \\
&=& \left[ (i m_\psi) \left( g^{tt} (i \omega^*) +g^{t \psi} (-i m_\psi)  \right) + (-i m_\psi)  \left( g^{tt}(-i \omega) +g^{t \psi} (i m_\psi) \right) \right] \Psi \Psi^*  \nonumber \\
&=&-2  m_\psi \left[  \left( g^{tt} \omega_R - g^{t \psi}  m_\psi \right)  \right] \Psi \Psi^*
\eea
so
\bea
L &=& -  2 m_\psi  \int  \sqrt{-g} dr d \angles ~   \left( g^{tt} \omega_R - g^{t \psi}  m_\psi \right)  \Psi \Psi^* \label{Def:AngMom}
\eea
The integral extends over the entire spacelike slice. We will however look at the contributions from the inner and outer regions separately in the next subsection. 

We next find the energy. The stress energy tensor gives us
\bea
T_t^{\p t} &=& \partial_t \Psi \partial^t \Psi^* + \partial^t \Psi \partial_t \Psi^* -\h \left( \partial_t \Psi \partial^t \Psi^* + \partial^t \Psi \partial_t \Psi^*+ \partial_i \Psi \partial^i  \Psi^* +\partial^i \Psi \partial_i  \Psi^* \right) \nonumber  \\
&=& \h \left[ \partial_t \Psi \partial^t \Psi^* + \partial^t \Psi \partial_t \Psi^*- \partial_i \Psi \partial^i  \Psi^* -\partial^i \Psi \partial_i  \Psi^*\right] \nonumber \\
&=& g^{tt} \partial_t \Psi \partial_t \Psi^* - g^{ij} \partial_i \Psi \partial_j \Psi^*
\eea
where it should be observed that in the last step all terms with single derivatives in time canceled. Thus the energy is
\bea
H&=& - \int  \sqrt{-g} dr d \angles ~ \left[ g^{tt} \partial_t \Psi \partial_t \Psi^* - g^{ij} \partial_i \Psi \partial_j \Psi^*\right] 
\eea
Using integration by parts we can write
\be
H=H_{bulk} + H_{boundary}
\ee
where
\bea
H_{bulk}&=&- \int \sqrt{-g}  dr d \angles ~ \left[  g^{tt} \partial_t \Psi \partial_t \Psi^* \right] \nonumber \\
&&- \h  \int   dr d \angles ~ \partial_i \left[ \sqrt{-g} g^{ij}  \partial_j \Psi \right] \Psi^* - \h  \int   dr d \angles ~ \partial_i \left[ \sqrt{-g} g^{ij} \partial_j \Psi^* \right]  \Psi
\eea
and
\be
H_{boundary}=\h \int   dr d \angles ~ \partial_i \left[ \sqrt{-g} g^{ij} \left( \partial_j \Psi \Psi^* + \partial_j \Psi^* \Psi \right) \right]  \label{Def:EnergyBoundary}
\ee
Here we imagine that $H_{bulk}$ is being carried out over some region of the spacelike slice, and $H_{boundary}$ can be written as an integral  over the boundary of this region. 

Using the equation of motion
\be
\partial_\mu(\sqrt{-g} \partial^\mu \Psi)= \partial_t(\sqrt{-g} \partial^t \Psi) + \partial_i(\sqrt{-g} g^{it} \partial_t \Psi)+ \partial_i(\sqrt{-g} g^{ij} \partial_j \Psi)=0
\ee
we get 
\bea
H_{bulk}&=&- \int \sqrt{-g}  dr d \angles ~ \left[  g^{tt} \partial_t \Psi \partial_t \Psi^* \right] \nonumber \\
&&+\h \int   dr d \angles ~ \left[ \partial_t \left( \sqrt{-g}   \partial^t \Psi \right) \Psi^* + \partial_i \left( \sqrt{-g}  g^{it} \partial_t \Psi \right) \Psi^*  \right ] \nonumber \\
&&+\h \int   dr d \angles ~ \left[ \partial_t \left( \sqrt{-g}   \partial^t \Psi^* \right) \Psi + \partial_i \left( \sqrt{-g}  g^{it} \partial_t \Psi^* \right) \Psi  \right]
\eea
The only terms that are of the form $g^{\mu t}$ are $g^{tt},g^{t\psi}, g^{t \phi}$ and $g^{ty}$, so the index $i$ in the above equation can only range over $t, \psi, \phi, y$. Since the metric doesn't depend on $t, \psi,\phi,y$ we can pull out the metric from the derivatives. Thus
\bea
H_{bulk}&=& -\int \sqrt{-g}  dr d \angles ~ \left[  g^{tt} \partial_t \Psi \partial_t \Psi^* \right] \nonumber \\
&&+\h \int  \sqrt{-g} dr d \angles ~ \left[ \left( \partial_t \partial^t \Psi \right) \Psi^* +  g^{it} \left( \partial_i \partial_t \Psi \right) \Psi^*  \right ] \nonumber \\
&&+\h \int  \sqrt{-g} dr d \angles ~ \left [ \left( \partial_t \partial^t \Psi^* \right) \Psi + g^{it} \left( \partial_i \partial_t \Psi^* \right) \Psi  \right \}
\eea
which can be rewritten as
\bea
H_{bulk}&=&- \int \sqrt{-g}  dr d \angles ~ \left[  g^{tt} \partial_t \Psi \partial_t \Psi^* \right] \nonumber \\
&&+\h \int  \sqrt{-g} dr d \angles ~ \left[ g^{tt} \left (  \partial_t \partial_t \Psi \right) \Psi^* + 2 g^{it} \left( \partial_i \partial_t \Psi \right) \Psi^*  \right ] \nonumber \\ 
&&+\h \int  \sqrt{-g} dr d \angles ~ \left [g^{tt}  \left( \partial_t \partial_t \Psi^* \right) \Psi + 2 g^{it} \left(  \partial_i \partial_t \Psi^* \right) \Psi  \right \} 
\eea
Using  the ansatz \bref{Eqn:Ansatz} in this expression we get
\bea
H_{bulk}&=& -\int \sqrt{-g}  dr d \angles ~ \Bigg [ g^{tt} \Big \{ (-i \omega) (i \omega^*)-\h (-i\omega)^2 - \h(i\omega^*)^2  \Big\} \nonumber \\
&&~~~~~~~~~~~~~~~~~~~~~~~~~~~~~~-g^{t \psi}  \Big \{ (i m_\psi)(-i \omega) + (-i m_\psi) (i\omega^*)  \Big\} \Bigg] \Psi \Psi^* \nonumber \\
\eea
which simplifies to
\bea
H_{bulk}&=& -2 \omega_R \int \sqrt{-g}  dr d \angles ~ \left[ g^{tt} \omega_R - g^{t \psi} m_\psi  \right] \Psi \Psi^*
\eea
We thus see that
\bea
H &=& L \f{\omega_R}{m_\psi}   +H_{boundary} \label{Def:Energy}
\eea
where we used (\ref{Def:AngMom}). Thus we see that upto a boundary term, the integral involved in computing the angular momentum charge will also give us the integral required for computing the energy.

We will now compute the conserved quantities in the inner and outer regions.

\subsection{The angular momentum in the inner region}

The metric in the inner region is (\ref{metricinner}). Recall that $\rho={R\over \sqrt{Q_1Q_5}} r$, and we will use both the  variables $r$  and  $\rho$ in the equations below. ($r$ is a more natural coordinate at infinity, while $\rho$ simplifies the metric in the $AdS$ region.)

The inverse metric (with the coordinates ordered as  $t,y,r,\theta,\psi,\phi$) is
\be
g^{\mu \nu}= \f{1}{\sqrt{Q_1 Q_5}} \left(
\begin{array}{llllll}
 -\frac{R^2}{\rho ^2+1} & 0 & 0 & 0 & \frac{m R}{\rho ^2+1} & 0 \\
 0 & \frac{R^2}{\rho ^2} & 0 & 0 & 0 & -\frac{m R }{\rho ^2} \\
 0 & 0 & \f{Q_1 Q_5}{R^2} (\rho ^2+1) & 0 & 0 & 0 \\
 0 & 0 & 0 & 1 & 0 & 0 \\
 \frac{m R}{\rho ^2+1} & 0 & 0 & 0 & \sec ^2(\theta )-\frac{m^2}{\rho
   ^2+1} & 0 \\
 0 & -\frac{m R}{\rho ^2} & 0 & 0 & 0 & \frac{m^2}{\rho ^2}+\csc ^2(\theta
   )
\end{array}
\right) \label{Eqn:AdSInvMetric}
\ee
The determinant of the metric is
\be
g=-\f{(Q_1 Q_5)^2}{R^2} \rho^2 \cos^2 \theta \sin^2 \theta \label{Eqn:AdSDetMet}
\ee
From (\ref{Def:AdSCood}) and  \bref{Def:xBulk} we see that $\rho^2=x$, where $x$ is the radial variable which we had used in writing the wavefunction. From \bref{Eqn:WavefunctionIn} the wave function is thus
\be
\Psi = e^{-i (\omega t - m_\psi \psi)} \chi(\theta) (1+ \rho^2)^{-\f{l+2}{2}} 
\ee
So from \bref{Def:AngMom} the contribution to the angular momentum $L$ from  this region is
\be
L_{in} =  - m_\psi (4 \pi C \;  Q_1 Q_5 )  e^{2 \omega_I t}   \int_0^{(Q_1 Q_5)^\f{1}{4}} dr ~ \rho ~   \left( g^{tt} \omega_R- g^{t \psi}  m_\psi \right)(1+ \rho^2)^{-(l+2)} 
\ee
where
\be
C=\int d\Omega |\chi (\theta) |^2 
\ee
Here the $ 2 \pi R$ comes from integral over the $y$ direction and 
\be
d \Omega = \cos \theta \sin \theta d \theta d \phi d \psi
\ee
Using  the inverse metric (\ref{Eqn:AdSInvMetric}) we see that
\be
L_{in} =  m_\psi (4 \pi C \;  Q_1 Q_5 )   e^{2 \omega_I t}  \left(  \omega_R R +m  m_\psi \right)   \int_0^{ {R}/{(Q_1 Q_5)^\f{1}{4}}} d\rho ~ \rho ~      (1+ \rho^2)^{-(l+3)} 
\ee
The integrand goes to zero for large $\rho$. In the large $R$ limit (eq. (\ref{Def:epsilon})) the upper bound on $\rho$ is large, so to leading order in $\epsilon$ we can set the upper limit of the integral to infinity. This gives
\be
L_{in}=  m_\psi ( 2 \pi C \; Q_1 Q_5 )  e^{2 \omega_I t}   \left(  \f{\omega_R R +m  m_\psi}{l+2} \right) \label{Eqn:LAdS}
\ee
The quantum numbers of the waveform satisfy the relation  (eq. \bref{wtwo})
\be
\omega_R R +m  m_\psi = - (l+2) 
\ee
So we get for the contribution to angular momentum $L$ from the inner region
\be
L_{in}= -m_\psi \left( 2 \pi C \; Q_1 Q_5 \right)  e^{2 \omega_I t} 
\ee
\subsection{The angular momentum in the outer region}

In the outer region the metric was flat spacetime (\ref{flat}) to leading order.
We have $\sqrt{-g} = r^3 \cos \theta \sin \theta$. From the wavefunction \bref{Eqn:WavefunctionOut} and \bref{Def:AngMom} we get
\be
L_{out} = m_\psi  \left (4 \pi C ~ Q_1 Q_5\right)   \omega_I e^{2 \omega_I t}  \int_{(Q_1 Q_5)^\f{1}{4}}^\infty dr ~ e^{-2 \omega_I r}
\label{outl}
\ee
Which gives us to leading order
\bea
L_{out}&=& m_\psi \left( 2 \pi C \; Q_1 Q_5 \right)  e^{2 \omega_I t} 
\eea
We see that 
\be
L_{in}+L_{out}=0
\ee
as was required.

\subsection{The energy}

Now let us look at the contributions to the energy $H$ from the inner and outer regions. 

From \bref{Def:Energy} we  see that in the inner region the contribution to the energy is
\be
H_{in}=-\omega_R \left( 2 \pi C \; Q_1 Q_5 \right)  e^{2 \omega_I t} + H_{boundary}^{in}
\label{tinner}
\ee
and in the outer region it is
\be
H_{out}=\omega_R \left( 2 \pi C \; Q_1 Q_5 \right)  e^{2 \omega_I t} + H_{boundary}^{out}
\label{touter}
\ee
The boundary terms arise from integration over $r$. For the inner and outer  regions we get
\bea
H_{boundary}^{in}&=& \h \int d \angles ~ \left[ \sqrt{-g} g^{rr} \left( \partial_r \Psi \Psi^* + \partial_r \Psi^* \Psi \right) \right] |_{r=(Q_1 Q_5)^\f{1}{4}} \nonumber \\
&&-\h \int d \angles ~ \left[ \sqrt{-g} g^{rr} \left( \partial_r \Psi \Psi^* + \partial_r \Psi^* \Psi \right) \right] |_{r=0} \nonumber \\
&\equiv&H^{in, neck}_{boundary}+H^{in, r=0}_{boundary}
\eea
\bea
H_{boundary}^{out}&=&\h \int d \angles ~ \left[ \sqrt{-g} g^{rr} \left( \partial_r \Psi \Psi^* + \partial_r \Psi^* \Psi \right) \right] |_{r=\infty} \nonumber \\
&&- \h \int d \angles ~ \left[ \sqrt{-g} g^{rr} \left( \partial_r \Psi \Psi^* + \partial_r \Psi^* \Psi \right) \right] |_{r=(Q_1 Q_5)^\f{1}{4}} \nonumber \\
&\equiv&H^{out, r=\infty}_{boundary}+H^{in, neck}_{boundary}
\eea
Note that
\be
H^{in, neck}_{boundary}=-H^{out, neck}_{boundary}\equiv H^{neck}_{boundary}
\ee
 Using \bref{Def:xBulk} and \bref{Eqn:WavefunctionIn} we have for the inner region
\be
(\partial_r \Psi) \Psi^* = \partial_r x \partial_x \Psi ~ \Psi^*=  e^{2 \omega_I t} 2 r \f{R^2}{Q_1 Q_5} |\chi(\theta)|^2  \left( - \f{l+2}{2} \right) \f{1}{(1+x)^{l+3}}
\ee 
We get $H_{boundary}^{in, r=0}=0$ because of the factor $r$ in the above expression.  We get $H_{boundary}^{out, r=\infty}=0$  because $\Psi, \partial_r\Psi$ both vanish as $\sim e^{-\omega_I r}$ at $r\rightarrow\infty$.

We now note that the terms at the neck $H_{boundary}^{neck}$ are subleading compared to the bulk terms in the inner and outer regions.  
To evaluate the term at the neck we observe that at the neck $r \sim (Q_1 Q_5)^\f{1}{4}$. Recalling the definition  (\ref{Def:epsilon}), we find that  $x \sim \epsilon^{-1}$. Let us use the inner wavefunction to estimate $H_{boundary}^{neck}$. We get
\be
(\partial_r \Psi) \Psi^* \sim e^{2 \omega_I t}  \f{1}{(Q_1 Q_5)^\f{1}{4}} \epsilon^{l+2}
\label{boundaryapprox}
\ee
From \bref{Eqn:AdSInvMetric} we get for  $g^{rr}$ at the neck
\be
g^{rr} \sim O(1)
\ee
and from \bref{Eqn:AdSDetMet} we get at the neck
\be
\sqrt{-g} \sim (Q_1 Q_5)^\f{3}{4}
\ee
So we get for the boundary term at the neck
\be
H_{boundary}^{neck} \sim e^{2 \omega_I t} (Q_1 Q_5)^\f{3}{4} R  \f{1}{(Q_1 Q_5)^\f{1}{4}} \epsilon^{l+2} =  e^{2 \omega_I t} (Q_1 Q_5)^\f{3}{4} \epsilon^{l+{3\over 2}}
\ee
Since $\epsilon$ is small the boundary term from the neck is subleading to the bulk terms in (\ref{tinner}),(\ref{touter}). Thus we have
\bea
H_{in}&\approx&-\omega_R \left( 2 \pi C \; Q_1 Q_5 \right)  e^{2 \omega_I t} \nonumber \\
H_{out}&\approx&\omega_R \left( 2 \pi C \; Q_1 Q_5 \right)  e^{2 \omega_I t} 
\eea

We see that
\be
H_{in}+H_{out}=0
\ee
as was required.

\subsection{Summary}

Let us summarize the computations of this section. The general structure of instabilities due to ergoregions is known \cite{kang,friedman,cominsschutz,ashtekar}. The instability can be modeled by a simple 1-dimensional problem exhibiting the `Schiff-Snyder-Weinberg effect' and the related `Klein paradox' \cite{schiff,klein,fulling}. When we quantize a field like the scalar field satisfying a second order differential equation, then we have a conserved norm that is not positive definite. For simple situations like the scalar field in flat spacetime, we can separate the solution into positive norm modes and negative norm modes, and associate  annihilation and creation operators respectively with these modes to quantize the field.  But in the presence of sufficiently strong potentials (electromagnetic, gravitational etc.) there can be a situation where we also have states with {\it zero} norm. These zero norm modes are paired up by the inner product, and can be quantized, but do not lead to a particle interpretation. Instead of the usual harmonic oscillator associated to each particle mode, we get a harmonic oscillator with an `upside down' potential. While the excitations of the usual harmonic oscillator gave particles, now we just have an instability that can grow with time. The unstable modes found in \cite{myers} are such zero norm modes.

The zero norm of such states arises from a cancellation of contributions from regions where the norm is positive and where it is negative. The vanishing charges arise in a similar way: there are regions with positive contribution and regions with negative contribution. For our problem, the part of the wavefunction at infinity has positive contribution to the norm and the energy, and a contribution to the angular momentum. The part in the ergoregion has negative norm, negative energy, and the opposite contribution to angular momentum. This is the case when we define norm and energy using the definitions natural at spatial infinity. An observer living in the ergoregion would use a different definition of local energy, with respect to which the energy of the wavefunction there would be positive. Thus in both the outer and inner regions physics is `normal': the allowed excitations have positive energy when we set up a definition of energy locally. But the definition of positive energy in these two regions is different, and so we are able to create particle pairs where the overall energy is zero (as seen from infinity).

More explicitly, the wavefunction at infinity has the form
\be
\Psi\sim e^{-i\omega t +i m_\psi \psi}
\ee
We have $\omega>0$, as expected for real quanta at infinity. In the inner region, we rewrite this wavefunction in the coordinates suited to the local geometry (\ref{metricinner}) there. In this geometry the coordinate $t$ is mixed with $\psi$ since  $g_{t\psi}\ne 0$.  Thus we move to new coordinates which do not have such a mixing. Using the rescaled coordinates  (\ref{Def:AdSCood}), we see that we can `unmix' the coordinates with the definitions
\be
\tau'=\tau, ~~~\psi'=\psi+m\tau
\ee
We can quantize the system in these new coordinates where the time $\tau'$ is orthogonal to the other directions. But in these coordinates the wavefunction looks like
\be
\Psi\sim e^{-i(\omega R+mm_\psi)\tau'+i m_\psi \psi'}\equiv e^{-i\omega' R \tau' +i m'_\psi \psi'}
\ee
Using (\ref{wtwo}) we see that
\be
\omega' R=\omega R+m m_\psi =-(l+2), ~~~m'_\psi=m_\psi
\ee
Thus the energy of the mode looks  negative in the new coordinate system, so that we would consider this an {\it antiparticle} mode. Thus the norm that we would associate to this mode in the local description would be opposite in sign to the contribution to the norm which we get when we quantize the  system as a whole using the time coordinate $t$. This behavior is characteristic of ergoregions.

In appendix \ref{atwo} we find the ergoregion for our geometries, in the large $R$ limit which we have taken. In appendix \ref{athree} we show that the wavefunction in the inner region is localized in this ergoregion. This localization is expected from the general process of ergoregion emission, but it is still helpful to see it explicitly for our given case.

\section{Radiation from fuzzballs}
\setcounter{equation}{0}

In the above sections we have performed some computations that detailed the  nature of the radiation produced by the special microstates of \cite{ross}. In this section we comment on the physical  properties of this radiation, and how more general microstates might be expected to behave.

\subsection{Bose enhancement}

In \cite{myers} it was shown that the microstates of \cite{ross} had an instability which led to emission of scalars at certain special frequencies. In \cite{chowdhurymathur} it was shown that this `instability emission' was exactly the `Hawking emission' that would be expected from these particular microstates. But the instability of \cite{myers} gave an exponential growth of a classical scalar field, while we normally think of Hawking emission as a slow  quantum process producing a thermal distribution of quanta. Thus these two processes at first appear to be quite different. How do we understand this apparent dissimilarity?

\begin{figure}[t] %  figure placement: here, top, bottom, or page
   \begin{center}\hspace{-1truecm}
   \includegraphics[width=1.6in]{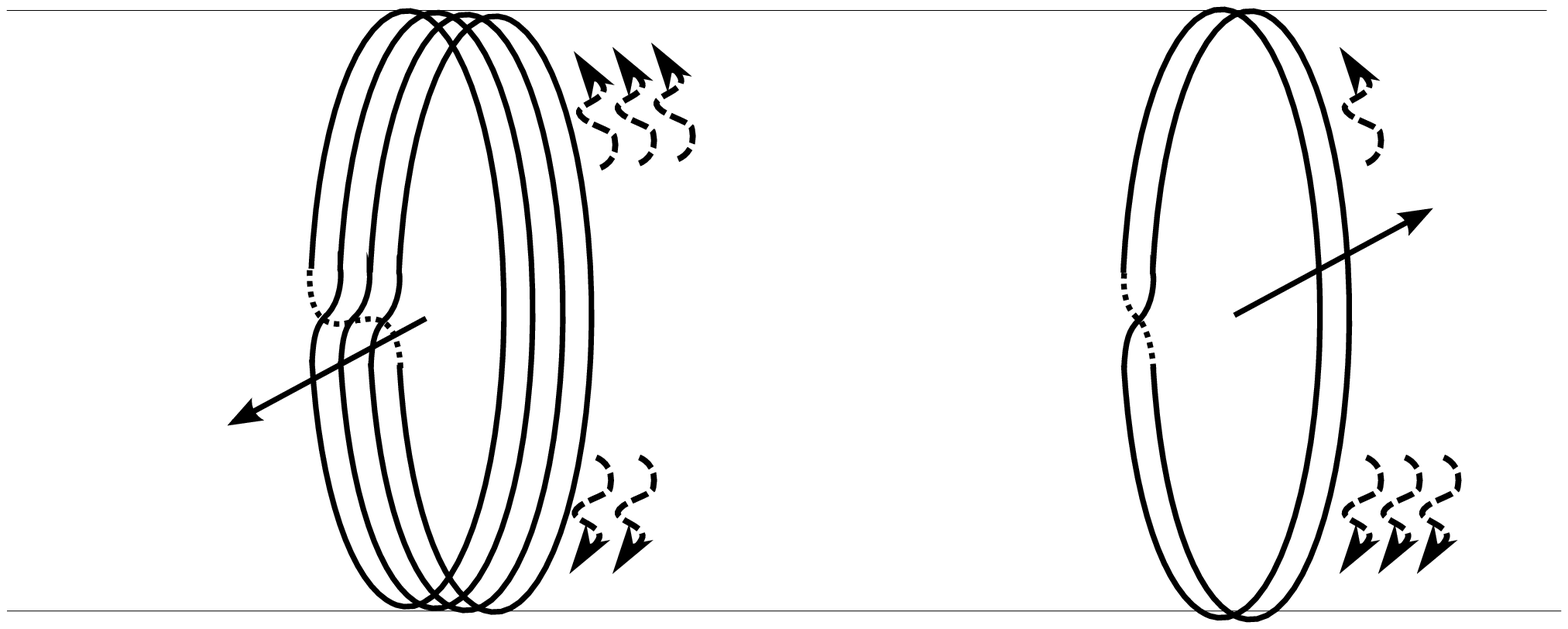}  \hspace{1truecm}
    \includegraphics[width=1.6in]{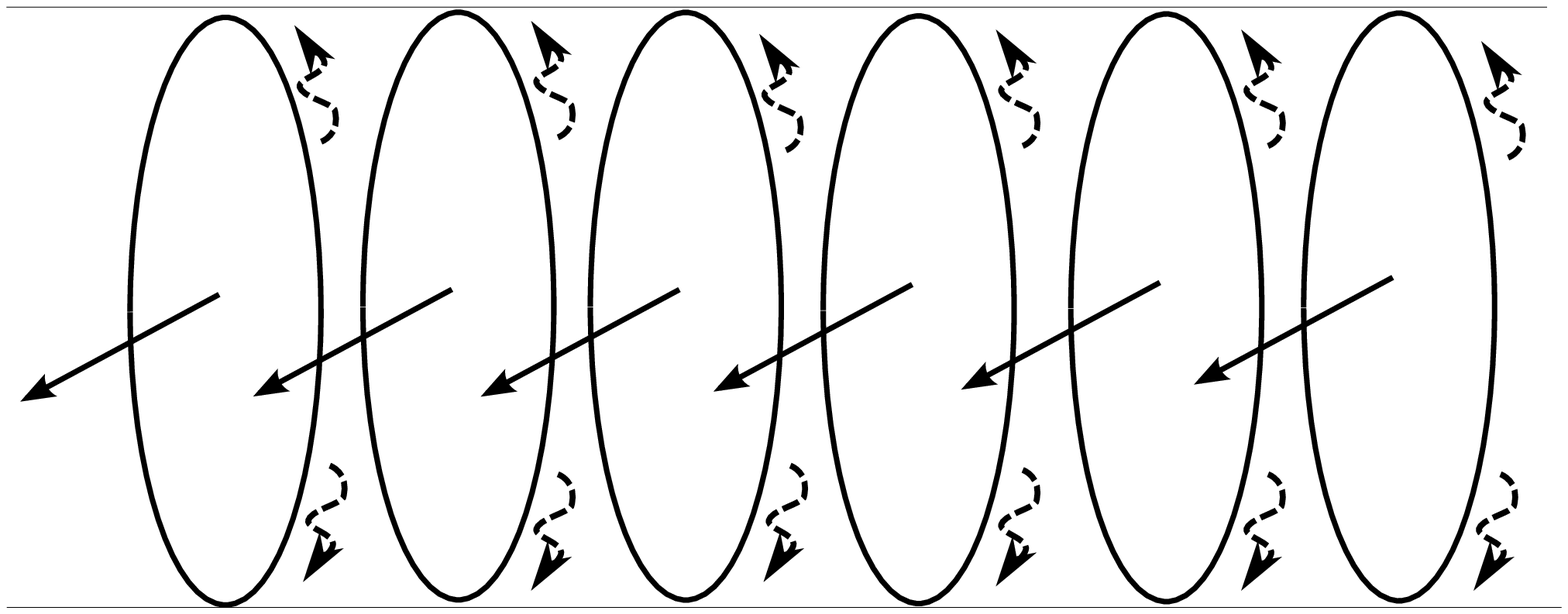}\hspace{1truecm}
    \includegraphics[width=1.6in]{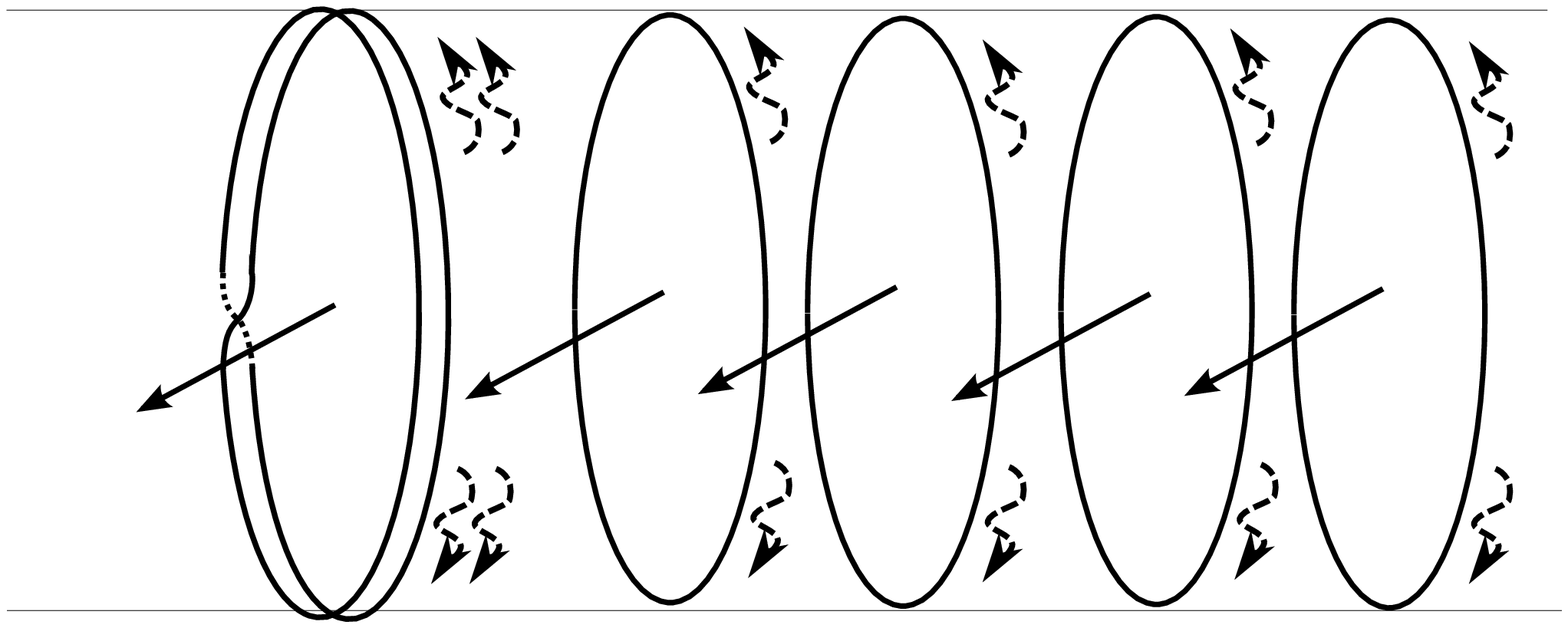} \\
    \vspace{.2truecm}
    \end{center}
    \hspace{2truecm} (a)\hspace{5truecm}(b) \hspace{4.2truecm}(c)
     \caption{(a) A generic CFT state (b) The special microstates of \cite{ross} (c) The CFT state after emission of a few quanta.}
   \label{fig:one}
\end{figure}

The answer lies in the special nature of the microstates that we are looking at. This issue was discussed in brief in \cite{chowdhurymathur}, but we look at it in more detail here.

First we need to recall the description of microstates in the CFT picture.
Our system had $n_1$ D1 branes wrapped in $S^1$, and $n_5$ D5 branes wrapped on $T^4\times S^1$. The bound state of these branes can be described by an `effective string' which has winding number $n_1n_5$ on the $S^1$.  These $n_1n_5$ units of winding can be decomposed into one or more `component strings', where the component string $i$  winds  $m_i$ times around the $S^1$ before closing.  Each component string carries a `base spin' in the representation $({1\over 2}, {1\over 2})$ of the rotation group in the noncompact directions $SO(4)\approx SU(2)\times SU(2)$. 

Different states of the extremal D1-D5 system are given by different ways of breaking the effective string into component strings, and by choosing different spins for these component strings. Excited states of the D1-D5 system are generated by exciting left and right moving excitations on the component strings. The allowed left excitations are 4 bosons $X^1, X^1, X^3, X^4$ and 4 fermions $\psi^1, \psi^1, \psi^3, \psi^4$. The allowed right moving excitations are 4 bosons $X^1, X^1, X^3, X^4$ and 4 fermions $\tilde\psi^1, \tilde\psi^1, \tilde\psi^3, \tilde\psi^4$.

In fig.\ref{fig:one}(a) we depict a generic CFT state of the nonextremal black hole. The component strings have large winding number, randomly oriented base spins, and a thermal distribution of left and right moving excitations $X, \psi, \tilde\psi$. In fig.\ref{fig:one}(b) we depict the CFT state for the special geometries that we have considered. All component strings are `singly wound', all base spins are aligned, and all left and right moving excitations are fermionic. Further, these fermionic excitations are taken to fill up the allowed energy levels upto a `fermi surface', and their spins are all aligned with the base spins of the component strings. Thus the state has maximal angular momentum for its energy.

Excited states of the CFT can emit quanta that leave the system as `Hawking radiation'. A quantum of angular momentum $l$ is emitted in a process where $l+1$ strands of the effective string are twisted together, and the excitations on the strands can be created or annihilated by appropriate operators. Thus each microstate of the system will emit differently from every other microstate. When this process is applied to the generic state of fig.\ref{fig:one}(a), the emitted radiation rate is found to agree with the semiclassical Hawking radiation from the corresponding near-extremal D1-D5 black hole. When we apply the same emission process to the  specific  microstate of fig.\ref{fig:one}(b), then we will get the emission particular to this microstate, and since this microstate is quite different from the generic one, the emission will also look different from the semiclassical Hawking radiation spectrum. Nevertheless, whatever emission we get will be the `Hawking radiation' for this specific microstate.

In fig.\ref{fig:one}(c) we depict the twisted strings produced in the emission process, where the starting state was the special microstate of fig.\ref{fig:one}(b). Let us denote the number of initial component strings by
\be
N\equiv n_1n_5
\ee
After $n$ quanta have been emitted, we have $n$ twisted strings, and 
\be
N_u\equiv N-(l+1)n
\ee
untwisted strings. Consider the emission of the next quantum from the system. All the untwisted strings are identical, so they behave like bosons. The twisted strings are also identical, so they also behave like bosons. Thus when we annihilate $l+1$ untwisted strings from the system and create one twisted string, we get the `bose factors'
\be
\sqrt{N_u}\sqrt{N_u-1}\dots \sqrt{N_u-l}\sqrt{n+1}
\ee
in the transition amplitude ${\cal A}$ \cite{chowdhurymathur}. (The emitted quantum is also a boson, but since it escapes to infinity, we do not get a bose enhancement factor for it.)

The probability for emission, and thus the rate of emission $\Gamma$ then has a factor
\be
\Gamma\sim |{\cal A}|^2\propto N_u(N_u-1)\dots (N_u-l) (n+1)
\label{bose}
\ee
At the start of the emission process, $n$ is small, so we can write
\be
N_u\approx N,~~N_u-1\approx N, ~~\dots~~N_u-l\approx N
\label{approx}
\ee
and we find
\be
\Gamma\propto (n+1)
\ee
In this factor $(n+1)$ the $1$ gives spontaneous emission and the $n$ gives stimulated emission. In the classical limit where several quanta have been emitted we have $n\gg 1$, and  we can write
\be
n+1\approx n
\ee
Then we get
\be
\Gamma={dn\over dt}\propto n
\label{exp}
\ee
so that the emission grows exponentially. In \cite{chowdhurymathur} it was shown that the frequency of emission and the rate of exponential growth agree exactly between the instability seen in gravity and the emission computed from the CFT.

Let us now come to our question: why do we get an exponential growth for our specific microstate, when the generic one gives the slow thermal emission expected of Hawking radiation? Of course each microstate will emit in a way that is different from any other microstate. But what makes our paricular microstates  special is that all component strings are `in the same state'. This has two consequences, both of which can be seen in (\ref{bose}). One is that the twisted strings which are created are all created in the same state, and so successive twisted strings have a bose enhanced probability of production; this gives the exponential growth (\ref{exp}). The other is that there is a large supply of component strings in the initial state, so that we have been able to use the approximation (\ref{approx}). 

The gravity computation of \cite{myers} follows the emission only to the point where the perturbation stays linear; this means that the backreaction of the perturbation on the geometry can be ignored. We have seen in the present paper that the emission process deposits quanta in the ergoregion in a specific wavefunction. Thus after $n$ quanta have been deposited, there will be a bose enhancement factor  $\approx n$ for the deposition of the next quantum, and we can understand the exponential growth in the gravity picture just like we understand it in the CFT picture. But note that when $n$ becomes sufficiently large then we will no longer be able to ignore the backreaction of these quanta on the geometry. At that point the linear perturbation approximation used in \cite{myers} will fail, and we will have to solve the full gravity problem. On the CFT side, this would correspond to $n$ becoming large enough that the approximation (\ref{approx}) cannot be made any more. We see from the CFT computation that when $n$ becomes this large then the emission will slow down and eventually stop, since we will run out of the component strings in the initial state.

Now consider a generic state. In a generic state we will have only a few component strings of any type, and the same will hold for the component strings after emission. Thus there is {\it no} bose enhancement (\ref{bose}), and also no large supply of initial state strings for a given emission mode. Thus only a few quanta of any given type will be produced, and thus no classical instability will be seen. As we know from the computations of \cite{radiation}, the emission will now agree with the semiclassical Hawking rate.

To summarize, the same CFT emission vertex that reproduces Hawking emission from the generic CFT state gives the instability radiation from our specific microstate. Thus the instability radiation must be considered the Hawking radiation from our special microstates. The situation is somewhat similar to the relation between laser emission and blackbody radiation. In a laser all quanta are in the same mode, and we can describe the physics by classical values of the electromagnetic fields. As we distribute the quanta over more and more modes, the state becomes more `quantum', and finally when the occupation number of the typical mode reached $O(1)$, we arrive at the generic state of blackbody radiation. Similarly, we can imagine starting with out special CFT state where all component strings are in the same state, then moving to a state where the component strings are of a few different types, and so on all the way to a generic state where the component strings are highly twisted, with only $O(1)$ number of excitations of any given energy. At this last step we will reach the generic state, but the gravity state would have become progressively more complicated in the process, ending up as a quantum `fuzzball'.

\subsection{Generic ergoregions}

The emission process noted in \cite{myers} was the process of `ergoregion emission'. The geometry has a Killing vector, but while this Killing vector is timelike at infinity, it is not timelike everywhere; it becomes spacelike inside the ergoregion. Thus we cannot set up a time-independent vacuum for the system using the Killing vector, and there will be pair production out of the vacuum for such a geometry. (See \cite{ashtekar} for a general discussion of pair production in ergoregions.)

Our special geometries have high rotation along the directions specified by their axial symmetry, and in such situations the Killing vector becomes spacelike by acquiring a sufficiently large component in the direction of rotation. But we do not need to have any such axial symmetry in order to have an ergoregion. 

Consider a star cluster; this is a group of stars that are orbiting around in the mean gravitational field that they produce. Let each star be rotating  around its axis fast enough to have an ergoregion, but let the rotation axes of the different stars be oriented at random, so that the star cluster has no net rotation. What will an observer outside the cluster see? There will presumably be radiation from each star due to particle production in its ergoregion, and so the cluster will as a whole radiate quanta to infinity. But the geometry has no axial symmetry and no net rotation. 

We can imagine that a similar situation will hold for a large class of black hole microstates. Starting with the axially symmetric geometry that we have studied, we can imagine making a small deformation that destroys the axial symmetry; this will not remove the ergoregion and thus the geometry will still radiate quanta. In fact the only property that we need from the geometry is that there be no timelike Killing vector. In the axially symmetric geometry all timelike observer worldlines inside the ergoregion had to rotate in the $\psi$ direction. More generally the light cones can tilt in such a way that observer worldlines inside the ergoregion are forced to move in {\it some} direction $\zeta^i(x)$. Then we cannot set up a time independent vacuum and we will create particle pairs; radiation of these pairs can be some or all of the Hawking radiation from the microstate.

To summarize, in the traditional picture of Hawking emission we do not have a global timelike Killing vector, but this happens because the norm of Killing vector  $\partial_t$ vanishes at the horizon and becomes spacelike inside the horizon. In our special family of microstates, there is again no global timelike Killing vector, but instead of a horizon there is an ergoregion; the norm of the Killing vector vanishes at the boundary of this ergoregion and the Killing vector becomes spacelike inside the ergoregion. There is particle production in both cases, since there is no time independent vacuum in either case. Different microstates can have ergoregions of different shapes, and these ergoregions can be very complicated, with no particular symmetry in the overall geometry. 

For generic states the ergoregions can be very `shallow'; i.e., after only a few quanta collect in them, the backreaction becomes order unity, and the emission shuts off. This would correspond, in the CFT picture, to having only a few excitations of the given type: when these are emitted, there is no emission in that given mode, and there is thus no exponential build up of a classical perturbation. Thus it is  possible that very complicated ergoregions produce an emission which has the spectrum of the semiclassical Hawking emission.

Note that `stars' with ergoregions are  different from black holes with ergoregions. In the case of black holes, a particle pair can be created in the ergoregion, one member flows off to infinity and the other falls through the horizon \cite{myersp}. Thus there is no `bose enhancement' and no exponential growth of the perturbation. In a star a similar pair creation can occur, but the quanta falling into  the star collect in the star and lead to a bose enhancement which gives an exponential growth of the perturbation. Our general fuzzball states will be like stars in the sense that there will be no horizon, but we will avoid the exponential growth of the perturbation for a different reason: the ergoregions will be very complicated and `shallow', as explained above. Very complicated fuzzball states behave like black holes for practical purposes: for example a quantum that falls onto the fuzzball gets trapped and cannot emerge for long times, thus mimicking a fall through a horizon \cite{lm5}. Thus we would expect that a complicated fuzzball will reproduce the physics seen for black holes, where there is no exponential growth of the perturbation. It should be noted  that with complicated ergoregions $\omega_I$ can be very small, so the perturbation will appear to grow very slowly. Secondly, because the ergoregions will be `shallow',   a few quanta of backreaction will switch off production from any given mode in the ergoregion. Thus the exponential growth should not be seen for generic fuzzball states. 

\subsection{Different modes of emission}

In the CFT description there is essentially only one kind of emission: excited states of component strings fall to lower energy states and quanta are emitted. But when we work around a given background, it may be possible to divide up the overall emission into different types of processes. For our special set of microstates we can imagine two different kinds of emission:

(a) The ergoregion emission discussed above. Pairs are created out of the vacuum, one member escapes to infinity as radiation, while the other settles down in the ergoregion. For our special microstates, the emission spectrum is \cite{chowdhurymathur}
\be
\omega=-l-2-m_\psi m
\ee

(b) We can start with some quanta in the throat of the geometry of \cite{ross}. As discussed in \cite{hottube}, such quanta will bounce up and down the `capped' throat several times, with  a small probability of exiting the throat and escaping to infinity at the end of each bounce. This slow leakage from the throat can be another component of radiation from the state. For our special microstates this corresponds to starting with some component strings that have winding number $(l+1)$ and certain excitations; the emission process will then untwist these to $l+1$ singly wound component strings, and emit a quantum in the process. The spectrum for such emission can be computed to be 
\be
\omega=l+2-m_\psi m
\ee
The details of this computation will be presented elsewhere, but we note for now that this emission is obtained by solving the same perturbation equation as the one solved in appendix \ref{aone}, but choosing the solution that is exponentially growing at infinity instead of exponentially decaying at infinity. Such a solution is analogous to the one that is studied for $\alpha$-decay from a nucleus \cite{gamow}, and indeed the emission of quanta placed in the throat is similar to the escape of particles from the nuclear potential well.

\section{Discussion}
\setcounter{equation}{0}

We have done a few different things in this paper. First, we analyzed the ergoregion emission found in \cite{myers} in detail. We looked at the limit where the the radius $R$ of the $S^1$ was large, as this allowed a clean separation of the geometry into an `inner; and an `outer' region. We found that the wavefunction of the emitted scalar field could be split into two parts: one which gave quanta flowing off to infinity, and one which gave quanta settling down in the inner region. We showed that these two types of quanta had equal and opposite angular momenta and energy, as would be expected from a process of pair creation. In the appendices we showed that the quanta that settled in the inner region were localized inside the ergoregion. 

We then discussed how this particular example of radiation could extend to more general black hole microstates. 
The traditional picture of the black hole has a horizon, and Hawking radiation emerges from pair creation at this horizon. But with such a source for the radiation, we get information loss. String theory seems to tell us that equilibrium states of black holes are actually horizon sized quantum fuzzballs, so we do not have the traditional horizon which carried no information in its vicinity. Thus we can in principle get the information to emerge from the black hole microstate, but can we understand something more about the nature of the emission process?

For the special class of microstates constructed in \cite{ross}, we have seen that the emission arises due to the existence of an ergoregion, not due to a horizon. Ergoregions and horizons share the feature that in either case there is no time independent slicing of the geometry, so we cannot set up a time-independent vacuum state, and thus there will be particle production in general. But horizons would have led to information loss, and ergoregions (without horizons) do not. We conjectured that more general microstates would have more complicated ergoregions. We explained why the special class of microstates of \cite{ross} had a classical instability which caused emission at specific frequencies, while the traditional radiation from black holes is supposed to be a slow emission with thermal spectrum. This difference could be directly traced to the description in the dual CFT of the special states: they had all component strings `in the same state', so a phenomenon of bose enhancement made the emission strong and peaked at definite frequencies. The same emission computation applied to  the generic CFT state reproduces exactly the semiclassically expected spectrum of Hawking radiation, so this ergoregion emission is indeed just the `Hawking radiation' expected from these special microstates.

It would be good to get a more general understanding of generic non-extremal microstates. Of course the generic state will be very quantum, but can we approach such generic states through a family of classical states? Such a limit through classical geometries  can be taken for 2-charge extremal states \cite{lm4}, and may be possible for 3-charge extremal states, though the answer in the latter case is not clear as yet. We understand very few non-extremal states, so the answer here is not clear either. But it is interesting to note that all classical nonextremal geometries made so far are either time-dependent or have ergoregions. Thus in either case there is no time-independent Killing vector. In either case we will have pair creation, which we expect will  be the Hawking radiation for that state, just as was found for the family of \cite{ross} in \cite{chowdhurymathur}. 

It can be shown that if we take 3+1 dimensions and assume that there {\it is} a  timelike Killing vector, then there cannot be a spherically symmetric star with size smaller than  ${9\over 4}M$; more compact objects must be the Schwarzschild black hole \cite{wald}.\footnote{We thank G. Horowitz and J. Polchinski for pointing out this result to us.}  It is interesting that we cannot apply this result to the known classical nonextremal fuzzball states: they have no spherical symmetry, and they do not have a timelike Killing vector. Of course the generic state will be very quantum, and it is unclear how classical theorems could be used in such a case.

\section*{Acknowledgments}
\setcounter{equation}{0}

We thank Abhay Ashtekar, Steve Avery, Jeremy Michelson, Rob Myers, Anastasios Taliotis and Ak\i{}n Wingerter for many helpful comments. 
This work was supported in part by DOE grant DE-FG02-91ER-40690.

\appendix
\section{Solving the wave equation by `matching'}
\label{aone}
\renewcommand{\theequation}{A.\arabic{equation}}
\setcounter{equation}{0}

In this appendix we solve the wave-equation by matching solutions in the inner and outer regions. This computation was carried out in \cite{myers} and reproduced in a slightly different way in \cite{chowdhurymathur}, but we perform the matching again here because we need not only the frequencies of the modes but also explicit forms for the wavefunction in the inner and outer regions.

\subsection{The wave equation}

 We want to solve the scalar wave equation
\be
\Box \Psi = \f{1}{\sqrt{-g}} \partial_\mu(\sqrt{-g} \partial^\mu \Psi) = 0
\ee
in the geometry (\ref{Eqn:metric}). We follow the method of \cite{myers}.
The geometry has no  momentum along $y$ and no  rotation along $\phi$, and for simplicity we will let the perturbation also have these properties 
\be
\Psi = e^{-i (\omega t - m_\psi \psi ) }\chi(\theta) h(r) \label{Eqn:Ansatzp}
\ee
The wave equation reduces to an angular part 
\be
\f{1}{\sin 2 \theta} \partial_\theta (\sin 2 \theta \partial_\theta \chi) + [\Lambda + \omega^2 a_1^2 \sin^2 \theta - \f{m_\psi^2}{\cos^2 \theta} ]\chi=0 \label{Eqn:Angular}
\ee
and a radial part
\bea
&&\f{1}{r} \partial_r  ( r (r^2 + a_1^2-M) \partial_r h)  \nonumber \\
&& + \Big[ - \Lambda + \omega^2 (r^2 + M s_1^2 + Ms_5^2 + M)  +(a_1^2-M) \f{ (\omega R \f{c_1 c_5}{s_1 s_5} + m m_\psi)^2}{r^2 + a_1^2 -M} \Big] h =0 \label{Eqn:Radial}
\label{xieq}
\eea

\subsection{The Wave Equation in the large R limit}

From the CFT analysis we know that we are looking for wavefunctions with frequency $\sim {\f{1}{R}}$. For large $R$ we have $\epsilon\ll 1$, where $\epsilon$ was defined in (\ref{Def:epsilon}). We keep terms only upto leading order in $\epsilon$. In the large R limit the angular equation to leading order involves the laplacian on $S^3$
\be
\f{1}{\sin 2 \theta} \partial_\theta (\sin 2 \theta \partial_\theta \chi) + [\tilde \Lambda_1  - \f{m_\psi^2}{\cos^2 \theta} ]\chi=0
\ee
where $\tilde\Lambda_1$ is a constant to leading order: $\tilde \Lambda_1=l(l+2) + O(\epsilon^2)$.  The radial equation then becomes
\bea
\f{1}{r} \partial_r \left[ r \left(r^2 + \f{Q_1 Q_5}{R^2}  \right) \partial_r h \right]  +[- \tilde \Lambda_2 + \omega^2 r^2 ] h  +  \f{Q_1 Q_5}{R^2}\f{\xi^2}{r^2 + \f{Q_1 Q_5}{R^2} } h=0
\eea
In the large $R$ limit we have $c_i\approx s_i$ to order $O(\epsilon)$, so we have written the last term in (\ref{xieq}) by defining the variable
\be
\xi \equiv \omega R + m_\psi m   \label{Def:xi}
\ee
Further, we have
\bea
\tilde \Lambda_2 &=& l(l+2)  - \omega^2 (Q_1 + Q_5)
\eea
The second term above  is $O(\epsilon)$. The correction arising from the $O(\epsilon^2)$ terms in $\tilde\Lambda_1$ have been ignored.  We write
\be
\tilde \Lambda_2 = \nu^2-1
\ee
which gives
\be
\nu= l+1 + O(\epsilon) 
\label{Def:NuAndO1}
\ee
Define  
\be
x= r^2 \f{R^2}{Q_1 Q_5}  \label{Def:x}
\ee
The radial equation then becomes
\be
4 \partial_x(x (1+x) \partial_x h) + \left[(1- \nu^2) + \kappa^2 x \right]h + \f{ \xi^2}{1+x } h =0
\ee
where 
\be
\kappa  \equiv \omega \f{\sqrt{Q_1 Q_5}}{R} \label{Def:Kappa}
\ee
We can now solve the wave equation by looking at its approximations  for small $x$ and  for large $x$.

\subsubsection{Inner Region: $0< x \ll \f{1}{\epsilon}$}

The radial equation  can be approximated as
\be
4 \partial_x(x (1+x) \partial_x h) + \left[(1- \nu^2)  \right]h + \f{ \xi^2}{1+x } h =0
\ee
We can solve this with the ansatz $h=(1+x)^\f{\xi}{2} w$ and the requirement of regularity at $x=0$ to get
\be
h=(1+x)^\f{\xi}{2} \p_2F_1\left(\h(1-\nu+ \xi ),\h(1+ \nu+ \xi ),1,-x \right) \label{Eqn:InnerSolutionHyperGeometric}
\ee
From the relation
\bea
\p_2 F_1 \left(a,b,1,-x \right) &=&x^{-a} \frac{\Gamma (b-a) 
   }{\Gamma (1-a) \Gamma (b)} \p_2F_1\left(a,a;a-b+1;-\frac{1}{x}\right) \nonumber \\
   &&~~~+x^{-b} \frac{\Gamma (a-b)}{\Gamma (a) \Gamma
   (1-b)}  \p_2F_1\left(b,b;-a+b+1;-\frac{1}{x}\right) \nonumber \\
\eea
and the property $\p_2F_1(a,b,c,0)=1$ we get the large $x$ behavior
\bea
h= \f{\Gamma(\nu )}{\Gamma \left( \h(1+ \nu+ \xi ) \right) \Gamma \left(\h(1+ \nu-  \xi) \right)} x^{-\h(1- \nu)}  + \f{\Gamma(-\nu)}{\Gamma \left(\h(1-\nu+ \xi) \right) \Gamma \left(\h(1-\nu-  \xi) \right)} x^{-\h(1+ \nu)} \nonumber \\
\label{Eqn:MatchingInner}
\eea

\subsubsection{Outer region: $x \gg  1$}

In this region the radial part of the wavefunction can be written as
\be
4 \partial_x(x^2 \partial_x h) +\kappa^2 x h+(1-\nu^2) h =0
\ee
The solution to this is
\be
h = \f{1}{\sqrt{x}} \left[ C_1 J_{\nu}( \kappa \sqrt{x}) + C_2 J_{-\nu} (\kappa \sqrt{x}) \right]
\ee
The  behavior of this solution for large $\kappa \sqrt{x}$ is given by
\be
h =\f{1}{x^\f{3}{4}} \sqrt{\f{1}{2 \pi \kappa}}  \left[ e^{i \kappa \sqrt{x}} e^{-i \f{\pi}{4}}  \left( C_1 e^{-i \nu \f{\pi}{2} } + C_2 e^{i\nu \f{\pi}{2}} \right)  + e^{-i \kappa \sqrt{x}} e^{i \f{\pi}{4}}  \left( C_1 e^{i \nu \f{\pi}{2} } + C_2 e^{-i \nu \f{\pi}{2} } \right)  \right] \label{Eqn:VLargeH}
\ee
Imposing the condition that there are no ingoing waves we get
\be
C_1 + C_2 e^{-i \pi \nu}=0 \label{Eqn:BCOutgoing}
\ee
The solution in the outer region is then
\be
h=\f{1}{\sqrt{x}} C_2 \left[ -e^{-i \pi \nu} J_{\nu} (\kappa \sqrt{x}) +J_{-\nu} (\kappa \sqrt{x} ) \right] 
\ee
The small $x$ expansion of this is 
\be
h = \f{C_2}{\sqrt{x}} \left[  -e^{-i \pi \nu}  \f{1}{\Gamma(1+\nu)} \left(\f{\kappa \sqrt{x}}{2} \right)^\nu+ \f{1}{\Gamma(1-\nu)} \left(\f{\kappa \sqrt{x}}{2} \right)^{-\nu} \right] \label{Eqn:MatchingOuter}
\ee

\subsubsection{Matching The Solutions}

The two solutions overlap in the region $1 \lesssim  x \lesssim \f{1}{\epsilon}$. Equating the expressions \bref{Eqn:MatchingInner} and \bref{Eqn:MatchingOuter} we get 
\be
-e^{-i \pi \nu} \f{\Gamma(1-\nu)}{\Gamma(1+\nu)} \left( \f{\kappa}{2} \right)^{2 \nu} = \f{ \Gamma(\nu)}{\Gamma(-\nu)} \f{ \Gamma(\h(1-\nu +\xi)) \Gamma(\h(1-\nu -\xi))}{ \Gamma(\h(1+\nu +\xi)) \Gamma(\h(1+\nu -\xi))} \label{Eqn:Mathcing}
\ee
The left hand side is very small. This equation was solved in \cite{myers} by perturbing around the poles of one of the gamma functions in the denominator of the RHS. Perturbing around the poles of the gamma function on the left leads to an instability while perturbing around the poles of the gamma function on the right leads to an exponential decay. 
Since we are interested here in studying the instability we perturb around the poles of the gamma function on the left. To leading order we make the right hand side vanish by
\be
1+ \nu+ \xi =-2N, \quad N \in \mathbb N
\label{matcheq}
\ee
For simplicity we restrict attention here to the case $N=0$. From the definitions \bref{Def:xi} and  \bref{Def:NuAndO1} this implies
\be
\omega  = \f{1}{R} \left(-l-2-m_\psi m  \right)
\label{omegareal}
\ee
We will see that there is a small imaginary part to $\omega$, while the above equation gives the  leading order term for the real part of $\omega$. Thus
\be
\omega_R  = \f{1}{R} \left(-l-2-m_\psi m  \right) \label{Eqn:OmegaReal}
\ee
To solve the matching condition at the next level of iteration, 
 we perturb the location of the pole by varying the argument of the { Gamma} function. From (\ref{matcheq}) we get
\be
\delta \omega  = - \f{1}{R} 2\delta N \label{Eqn:DeltaOmega}
\ee
The corrections to $\delta \omega R$ coming from corrections to $\nu$ and $\xi$ are order $\epsilon$. As it turns out $\delta N$ is higher order in $\epsilon$. However the corrections from the former sources are real. We will see that $\delta N$ will be complex and its imaginary part will give the leading order contribution to $\omega_I$. We put 
\be
1+ \nu + \xi = -2 \delta N \label{Def:DeltaN}
\ee
back in the equation \bref{Eqn:Mathcing}. We assume that $\delta N \ll \epsilon$,  drop it from all the { Gamma} functions except the one whose argument is near a pole, and then note that the assumption is consistent.  We get
\be
-e^{-i\pi\nu}  \f{\Gamma(1-\nu)}{\Gamma(1+\nu)} \left( \f{\kappa}{2} \right)^{2 \nu} = \f{ \Gamma(\nu)}{\Gamma(-\nu)} \f{ \Gamma(-\nu ) }{ \Gamma(-\delta N) \Gamma(1+ \nu )}
\ee
which simplifies to
\be
\delta N = e^{-i\pi\nu}  \f{ \pi}{\sin(\pi \nu) \Gamma(\nu)^2} \left( \f{\kappa}{2} \right)^{2 \nu} \label{Eqn:DeltaN}
\ee
This gives to leading order
\be
\mathcal Im (\delta N)  = - \f{\pi}{(l!)^2} \left( \f{\kappa}{2} \right)^{2 (l+1)} 
\ee
From \bref{Eqn:DeltaOmega} this gives the leading order value for the imaginary part of $\omega$ 
\be
\omega_I =  \f{1}{R} \f{2\pi}{[l!]^2} \left( \f{\kappa}{2} \right)^{2 (l+1)} \label{Eqn:OmegaImag}
\ee

We now find the value of $C_2$ in eq. (\ref{Eqn:MatchingOuter}) by matching the inner and outer solutions.  Comparing the first terms on the RHS in \bref{Eqn:MatchingInner} and \bref{Eqn:MatchingOuter} we get 
\be
-C_2 \f{e^{-i\pi \nu}}{\Gamma(1+ \nu)} \left(\f{\kappa}{2}\right)^\nu = \f{\Gamma(\nu)}{\Gamma(\h(1+\nu + \xi)) \Gamma(\h(1+ \nu - \xi))}
\ee
We use \bref{Def:DeltaN} to get
\be
-C_2 \f{e^{-i\pi \nu}}{\Gamma(1+ \nu)} \left(\f{\kappa}{2}\right)^\nu = \f{\Gamma(\nu)}{\Gamma(-\delta N) \Gamma(1+ \nu )}
\ee
This along with \bref{Eqn:DeltaN} gives
\be
C_2 = \f{\pi}{\sin (\pi \nu) \Gamma(\nu)}  \left(\f{\kappa}{2}\right)^\nu \label{Eqn:C2}
\ee

\subsection{The solution to the wave equation}

Now that we have matched the inner and outer solutions, we can write down these solutions in their final forms. We will write down the solution to the leading order using \bref{omegareal}. In the next subsection we will show that this is indeed justified since the subleading term is much smaller.

Using \bref{omegareal} with $N=0$ and  $\nu \approx l+1$ in \bref{Eqn:InnerSolutionHyperGeometric} we get for the radial part of the wave function in the inner region
\be
h=(1+x)^{-\f{l+2}{2}} \label{Eqn:InnerRadialBranch1}
\ee
We then get the wavefunction in the inner region (using \bref{Eqn:Ansatzp})

\be
\Psi_{in}=e^{ \omega_I t} e^{-i(\omega_R t - m_\psi \psi) }\chi(\theta) (1+x)^{-\f{l+2}{2}} \label{Eqn:WavefunctionInAppendix}
\ee
Now let us find the wavefunction in the outer region. 
We can substitute the value of $C_2$ from \bref{Eqn:C2} into \bref{Eqn:VLargeH} and use the condition \bref{Eqn:BCOutgoing} to get 
\be
h =\f{i}{x^\f{3}{4}} \sqrt{\f{2 \pi}{\kappa}} \f{1}{\Gamma(\nu)}  \left(\f{\kappa}{2}\right)^\nu e^{i \kappa \sqrt{x}} e^{-i (\nu \f{\pi}{2} + \f{\pi}{4}) }
\ee
In terms of the radial coordinate $r$  we get (using \bref{Eqn:Ansatzp}, \bref{Def:x}, \bref{Def:Kappa} and \bref{Eqn:OmegaImag} and $\nu \approx l+1$)
\be
\Psi_{out} =   (i)^{3l}  e^{- i \f{\pi}{4}} \sqrt{ \f{Q_1 Q_5}{R} } \sqrt{ \f{\omega_I}{\omega_R}}   \chi(\theta) \f{1}{r^\f{3}{2}}  e^{ \omega_I (t-r)} e^{-i(\omega_R( t-r) - m_\psi \psi) }  \label{Eqn:WavefunctionOutAppendix}
\ee

\subsection{The comparison of the two terms at the neck} \label{compare}

We found the radial part of the solution in the inner region is given by \bref{Eqn:InnerSolutionHyperGeometric}. At the neck, this solution can be approximated as (\ref{Eqn:MatchingInner}). 
When we impose our physical boundary condition (purely outgoing waves at infinity), then we find that the frequencies are such that the first term in (\ref{Eqn:MatchingInner}) is small compared to the second term. We can therefore ignore this first term compared to the second. Applying this approximation to the entire inner solution (not just its behavior at the neck), we get the function $h$ given in (\ref{Eqn:InnerRadialBranch1}).  We then use this $h$ to get the inner wavefunction $\Psi_{inner}$ in equation (\ref{Eqn:WavefunctionInAppendix}). Here we estimate the ratio of the dropped part of the wavefunction  to the part that is kept, to verify that we could indeed ignore the correction which we drop.

At the neck $x  \sim \epsilon^{-1}$. We also recall that the spectrum is given by \bref{Def:DeltaN}. From these we get
\bea
h &=& \f{\Gamma(\nu)}{\Gamma(- \delta N) \Gamma(1+ \nu) } \epsilon^{\f{1-\nu}{2}}+ \f{\Gamma(-\nu)}{\Gamma(- \nu) \Gamma(1) } \epsilon^{\f{1+\nu}{2}} \nonumber \\
&=& \f{- \delta N}{\nu} \epsilon^{\f{1-\nu}{2}} +  \epsilon^{\f{1+\nu}{2}}  \nonumber \\
&=&- e^{-i\pi\nu}  \f{ \pi}{ \nu \sin(\pi \nu) \Gamma(\nu)^2} \left( \f{\kappa}{2} \right)^{2 \nu} \epsilon^{\f{1-\nu}{2}} +  \epsilon^{\f{1+\nu}{2}}   
\eea
Where we have used \bref{Eqn:DeltaN}. Noting that to leading order $\kappa \sim O(\epsilon)$ and $\sin \pi \nu \sim O(\epsilon)$ we see that the first term is of the order $O(\epsilon^{\f{3 \nu-1}{2}})\sim O(\epsilon^{\f{3 l+2 }{2}})$ while the second term is of the order $O(\epsilon^\f{1+ \nu}{2})\sim O(\epsilon^{\f{l+2}{2}})$. Thus the second term dominates over the first term even at the `neck'. We can also check that the radial derivative $\partial_r$ on the second term dominates over the radial derivative of the first term. These facts
allow us to drop the first term in writing the leading order approximation (\ref{Eqn:WavefunctionInAppendix}), which is wavefunction that we will use in (\ref{boundaryapprox}). 

\section{The Ergoregion}\label{atwo}
\renewcommand{\theequation}{B.\arabic{equation}}
\setcounter{equation}{0}

An asymptotically flat spacetime has an ergoregion when there is a Killing vector, but we cannot choose the Killing vector to be timelike everywhere. The region where the Killing vector is forced to be spacelike is called the ergoregion.

The metric \bref{Eqn:metric} has an ergoregion  if $m>1$ \cite{ross}. To see this we look at the Killing vector
\be
\xi^\mu=\delta^\mu_t + a \delta_y^\mu
\ee
where $a$ is a constant.
Requiring that the Killing vector be timelike at $r=\infty$ gives
\be
a^2 <1
\ee
Let us go to the large $R$ limit which we have used in all our computations. Then the metric in the inner region is given by 
 \bref{metricinner}. We have
\be
|\xi|^2= \f{\sqrt{Q_1 Q_5}}{R^2} \left[ (-1 - \rho^2+ m^2 \cos^2 \theta) + a^2 (\rho^2 + m^2 \sin^2 \theta) \right]
\ee
There will be no ergoregion if the norm of the Killing vector is everywhere negative for all $|a|^2 <1$. This would happen if
\be
(-1-\rho^2 +m^2 \cos^2 \theta) + a^2 (\rho^2 + m^2 \sin^2 \theta) < 0
\ee
This condition can be written as
\be
\cos^2 \theta < \f{1}{m^2} \left( \f{1- a^2 m^2}{1-a^2}+ \rho^2 \right)
\ee
For this condition to hold for all $\theta, \rho$ we would need
\be
1 <  \f{1}{m^2} \f{1- a^2 m^2}{1-a^2}
\label{ineqq}
\ee
If for a given $m$, the RHS above is necessarily less than unity, then we will {\it have} to have an ergoregion. Let $m>1$. Now we check which value of $a$ gives the largest value for the RHS. We find that this largest value is achieved for $a=0$, where the value of the RHS is 
\be
{1\over m^2}
\ee
So for $m>1$ the RHS of (\ref{ineqq}) is less than unity for all $a$, and the inequality cannot be satisfied. Thus we cannot construct a everywhere timelike Killing vector, and there will be an ergoregion. 

Now let us find the ergoregion. The most compact region where all timelike Killing vectors at asymptotic infinity become spacelike is called the ergoregion. Based on the above discussion, we set $a=0$. The ergoregion  is then  the region where
\be
g_{tt} >0
\ee
In our large $R$ limit this region is inside the `inner region', and is given by (using \bref{metricinner})
\be
-1-\rho^2 +m^2 \cos^2 \theta > 0 \label{Eqn:ergoregion}
\ee

\section{Localization of the inner wave function}\label{athree}
\renewcommand{\theequation}{C.\arabic{equation}}
\setcounter{equation}{0}

We have seen that in the large $R$ limit the instability of \cite{myers} creates a wavefunction of a special kind: one part flows off to infinity, while the other stays in the AdS region. In this appendix we look at the question of where in the AdS the latter part is localized. 

A wavefunction
of course spreads over all space, but if we go to a limit where it describes a classical trajectory, then we can ask where this trajectory is localized. Thus we take the limit where $l$ is large
\be
l\gg 1
\ee
 This makes the wavelength small compared to the curvature radius of $AdS_3\times S^3$, and thus gives a classical particle  trajectory in $AdS_3\times S^3$.

The wavefunction \bref{Eqn:WavefunctionIn} is seen to be a product of a part that describes the radial dependence in the AdS, and a part that describes the angular dependence in the $S^3$. The radial part is (using  \bref{Def:AdSCood} and  \bref{Def:x} to relate $x$ to $\rho$)
\be
|\Psi_{inner}|^2 \sim \f{1}{(1+ \rho^2)^{(l+2)}}
\ee
Thus for large $l$ the inner wavefunction is confined  to 
\be
\rho\approx 0
\label{rhozero}
\ee
For the angular variables, it is convenient to use coordinates where there is no mixing between the variables on $S^3$ and the variables in the AdS.  Writing
\be
\tau'=\tau, \qquad \psi'=\psi+m \tau, \qquad \varphi'=\varphi,  \qquad \phi'= \phi+ m \varphi \label{Eqn:Unfibered}
\ee
we get the metric for the inner region
\be
ds^2 = \sqrt{Q_1 Q_5} \left[ -(1+ \rho^2) {d\tau'}^2 + \rho^2 d \varphi^2 + \f{d \rho^2}{1+ \rho^2}+ d\theta^2 + \cos^2 \theta d{\psi'}^2 + \sin^2 \theta d {\phi'}^2 \right]
\ee
In this form the spacetime is manifestly $AdS_3 \times S^3$.  On the sphere we have for a particle
\be
p_\theta^2 + \f{p_{\psi'}^2}{\cos^2 \theta} + \f{p_{\phi'}^2}{\sin^2 \theta} = j^2 = const
\label{wone}
\ee
As the metric is manifestly independent of $\psi'$ and $\phi'$ we have
the conserved quantities
\bea
p_{\psi'}&=&j_{\psi'} =const \nonumber \\
p_{\phi'}&=& j_{\phi'} =0
\eea
where we taken $j_{\phi'}$ to vanish since we have not considered motion in the $\phi'$ direction in this paper. From (\ref{wone}), using $p_{\phi'}=0$ and $p_\theta^2>0$ we find that the particle trajectory is confined to
\be
\cos^2 \theta \ge \f{j_{\psi'}^2}{j^2} \label{Eqn:CosMax}
\ee
For the quanta that are radiated we have the basic relation (\ref{wtwo})
\be
\omega_R={1\over R}(-l-2-m_\psi m)
\ee
with $\omega_R >0$. This requires $m_\psi <0$ and we can  write 
\be
|m_\psi| m > l+2 \approx l
\ee
So we see that
\be
\f{|m_\psi|}{l} > \f{1}{m}
\label{eqqq}
\ee
From the coordinate redefinitions \bref{Eqn:Unfibered} we can see that
\be
m_{\psi'} = m_\psi
\ee
From our analysis with charge conservation we have seen that the sign of $j_\psi$ is opposite to $m_\psi$. Further,  $j^2=l^2$ since $j, l$ were obtained as separation constants arising from separation of angular variables in the laplace equation. From these relations between variables, (\ref{eqqq}) gives
\be
\f{j_\psi}{j} > \f{1}{m}
\ee
and together with \bref{Eqn:CosMax} we see that
\be
m^2 \cos^2 \theta > 1
\ee
From this relation and (\ref{rhozero}), and using  \bref{Eqn:ergoregion},  we see that the particle is localized inside the ergoregion.

\end{document}